\documentclass[prl,aps,amsfonts,amssymb,draft,floats,twocolumn,
showpacs]{revtex4}

\usepackage{graphicx}
\usepackage{epsf}

\newcommand{\be}{\begin{equation}}
\newcommand{\ee}{\end{equation}}

\begin{document}

\title{Ultrasharp Crossover from Quantum to Classical Decay in a Quantum Dot
Flanked by a Double-Barrier Tunneling Structure}
\author{Denis A. Gorokhov\footnote{e-mail: gorokhov@ccmr.cornell.edu}}
\affiliation{Laboratory of Atomic and Solid State Physics, Cornell University, Ithaca,
New York 14853-2501, U.S.A.}
\author{Rava A. da Silveira}
\affiliation{Lyman Laboratory of Physics, Harvard University, Cambridge, Massachusetts
02138, U.S.A. and Laboratoire de Physique Th\'eorique, Ecole Normale
Sup\'erieure, 24 rue Lhomond, 75005 Paris, France}

\begin{abstract}
The decay of metastable states is dominated by quantum tunneling at low
temperatures and by thermal activation at high temperatures. The escape rate
of a particle out of a square well is calculated within a semi-classical
approximation and exhibits an `ultrasharp' crossover: a kink in the decay
rate separates a \textit{purely quantum} regime at low temperatures from a 
\textit{purely thermal} regime at high temperatures. An experimental system
-- a quantum dot supplemented by a semiconductor heterostructure -- that may
be used to check the prediction, along with necessary experimental
conditions, are described.
\end{abstract}

\maketitle

\vskip1.5cm


The decay of metastable states\cite{Hanggi} is a phenomenon of great
generality, with realizations ranging from the creep of vortices in
superconductors\cite{Blatter} to the decay of false vacua\cite{Coleman} in
cosmology\cite{Linde}. A particular interest of the decay phenomenon resides
in the fact that it relates quantum to classical metastability: the decay
rate is highly sensitive to temperature, it is dominated by quantum
tunneling at low temperatures and by thermal activation at high temperatures.

\begin{figure}[t]
\epsfxsize= 0.8\hsize 
\epsffile{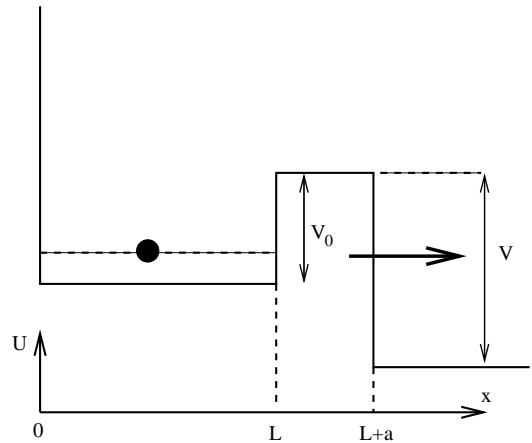} 
\caption{Illustration of our simple theoretical model, consisting of a
particle in a square well. Decay of the metastable state occurs by escape to
the right (represented by the bold arrow).}
\label{metastable_well}
\end{figure}

A canonical example consists of a quantum mechanical particle in an
asymmetric potential well, as illustrated in Fig.~\ref{metastable_well}.
Because of energetic metastability, sooner or later the particle leaves the
well and escapes to the right. The decay rate $\Gamma $, defined as the
inverse lifetime, depends on the form of the effective action at temperature 
$T$. At low $T$ the particle occupies its ground state most of the time and
escapes through quantum tunneling, so that $\Gamma \propto e^{-{S({0})/\hbar 
}}$, where $S({0})$ is the quantum mechanical action, while at high $T$ the
decay is Arrhenius-like, with $\Gamma \propto e^{-{V}_{0}/{T}}$.

Here, we calculate the decay rate of a particle initially residing in the
well of Fig. 1, and find that the transition from quantum to classical
behavior is `ultrasharp:' a singularity separates a purely quantum regime
from a purely classical regime. From a theoretical point of view, this
example is valuable as its semi-classical treatment is asymptotically exact
for large $V_{0}$. It also serves as a natural model to study an
experimentally realizable mesoscopic object, namely a quantum dot
supplemented by an adjacent double-barrier heterostructure. Below, we define
the object in question more precisely and discuss the specific experimental
conditions needed to observe an ultrasharp crossover.

Before deriving specific results, we briefly describe general properties of
the decay rate $\Gamma $. Mathematically, metastability may be encoded in
imaginary corrections to the energy levels $E_{n}= {\rm {Re}}%
E_{n}-i\hbar \Gamma _{n}/2$. Here ${\rm Re}E_{n}$ are the energy levels in
the large $V_{0}$ limit, in which the rates $\Gamma _{n}$ are small (limit
of `true metastability'). The probability $\mathcal{P}\propto {\left\vert
e^{-iE_{n}t/\hbar }\right\vert }^{2}$ that the system occupies a given state
then decays exponentially in time, according to $e^{-\Gamma _{n}t}$. If the
lifetimes $1/\Gamma _{n}$ are larger than the local thermal equilibrium
time, the initial preparation of the system is irrelevant; the particle
fluctuates in low-energy states according to the Boltzmann distribution.
Decay occurs because of rare fluctuations that drive the particle through
the energy barrier. In field-theoretic language, these fluctuations
correspond to imaginary time trajectories (instantons)\cite{Rajaraman} 
that come about as
solutions to saddle-point equations and, as a result, impose an exponential
dependence $\Gamma = A(T) e^{-S(T)/{\hbar }}$, 
where $S(T)$ is an effective
action. For true metastability $S(T)/\hbar \gg 1$ and
the main dependence of the decay rate upon temperature comes from the
action as, typically, the temperature dependence of the prefactor
$A(T)$ is weak.

\begin{figure}[t]
\epsfxsize= 0.8\hsize 
\epsffile{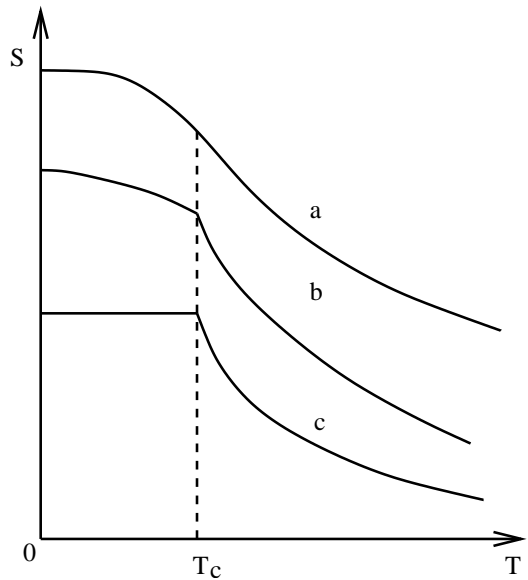} 
\caption{Different types of crossover from quantum to classical decay: (a)
smooth, (b) sharp, and (c) ultrasharp. At high $T$, thermal activation
dominates and $S(T) \propto1/T$ in all three curves.}
\label{different_possibilities}
\end{figure}

Generically, the action $S(T)$ varies from the ground state action $S(0)$ at 
$T=0$ to the high $T$ Arrhenius limit $\hbar V_{0}/T$, where $V_{0}\,$\ is
the height of the energy barrier to overcome. This crossover, nevertheless,
may occur in qualitatively different ways\cite%
{Affleck,Chudnovsky,Ch1,Gorokhov,MK,Kim,RLu}, depending on the shape of the
trapping potential and the metastable dynamics\cite{dynamics}, as
illustrated on Fig. \ref{different_possibilities}. For some metastable
systems, the function $S(T)$ is smooth within the whole temperature range,
but its second derivative is discontinuous at a critical temperature $T_{%
\text{c}}$ (curve (a) on Fig.~\ref{different_possibilities}). Phase
tunneling in a Josephson junction\cite{Larkin} constitutes a typical example
of such a behavior. It may also happen that the derivative of $S$ with
respect to $T$ has a discontinuity, resulting in a kink beyond which the
behavior becomes purely classical (with $S\propto 1/T$) (curve (b) on Fig.~%
\ref{different_possibilities}). Such a singular temperature dependence was
observed in $\mathrm{Mn}_{12}$ molecular magnets\cite{Kent}. Our example
yields yet another type of crossover, in which the quantum and classical
behaviors are completely separated (curve (c) on Fig.~\ref%
{different_possibilities}): as before the decay rate is purely classical
above the kink but, what is more, it is controlled by quantum tunneling 
\textit{only} (with a \textit{temperature-independent} action $S(0)$) below
the kink. Such ultrasharp transitions are easier to detect experimentally,
and might serve as useful tools for future investigations of macroscopic
quantum phenomena.

We emphasize that the curves in Fig. \ref{different_possibilities} are
qualitatively different. Curve (a) corresponds to the continuous deformation
of a given instanton as the temperature is increased. By contrast, curve (b)
results from the balance of two instantons, whose associated actions become
equal at $T_{\text{c}}$. We note that below $T_{\text{c}}$, thermal
fluctuations play a role as the effective action depends upon the
temperature. Curve (c) is a limiting case of curve (b), in which the minimal
action is the pure quantum action (corresponding to tunneling out of the
ground state) up to $T_{\text{c}}$. 

\bigskip

For the well of Fig. 1, the imaginary correction to the $n$th energy level $%
E_{n}$ is proportional to $\exp (-S_{E_{n}}/\hbar )$, with $S_{E_{n}}$ the
usual semi-classical (WKB) action 
\begin{equation}
S_{E_{n}}=2\sqrt{2m\left( V_{0}-E_{n}\right) }a.
\end{equation}%
If local thermal equilibrium is achieved fast enough, the particle in the
well occupies states that are very close to stable ($V_{0}\rightarrow \infty 
$) ones, with probabilities given by the Boltzmann weight. Thus, the decay
rate reads%
\begin{equation}
\Gamma \propto \sum_{n}\Gamma _{n}e^{-E_{n}/{T}},  \label{decay_rate}
\end{equation}%
with $\Gamma _{n}\propto e^{-S_{E_{n}}/\hbar }=e^{-2\sqrt{2m\left(
V_{0}-E_{n}\right) }a/\hbar }$. In the limit of true metastability, the sum
in Eq. (\ref{decay_rate}) runs over a large number of terms and is dominated
by the largest contribution, to wit the one that minimizes the function%
\begin{equation}
f(E)=\frac{S_{E}}{\hbar }+\frac{E}{T}.  \label{f_E}
\end{equation}%
It is easy to see that $f$ has no local minimum in $[0,V_{0}]$ as its second
derivative $\partial ^{2}f/\partial E^{2}$ is negative everywhere in the
interval. Consequently, $f(E)$ takes its smallest value either at $E=0$ (at
low $T$) or at $E=V_{0}$ (at high $T$). Precisely, 
\begin{equation}
\ln \left( \frac{1}{\Gamma }\right) \propto \frac{S(T)}{\hbar }=\left\{ 
\begin{array}{l@{\quad\quad}l}
{\frac{2\sqrt{2mV_{0}}a}{\hbar }}, & {T<\frac{\hbar }{2a}\sqrt{\frac{V_{0}}{%
2m}}\equiv T_{c},} \\ 
{\frac{V_{0}}{T}}, & {T>T_{c}.}%
\end{array}%
\right.  \label{final_result}
\end{equation}%
Hence decay results either from purely quantum tunneling or from thermal
activation out of the ground state. The ultrasharp crossover between the two
is signaled by a kink in $S(T)$, as in curve (c) of Fig. 2.

\begin{figure}[t]
\epsfxsize= 0.9\hsize 
\epsffile{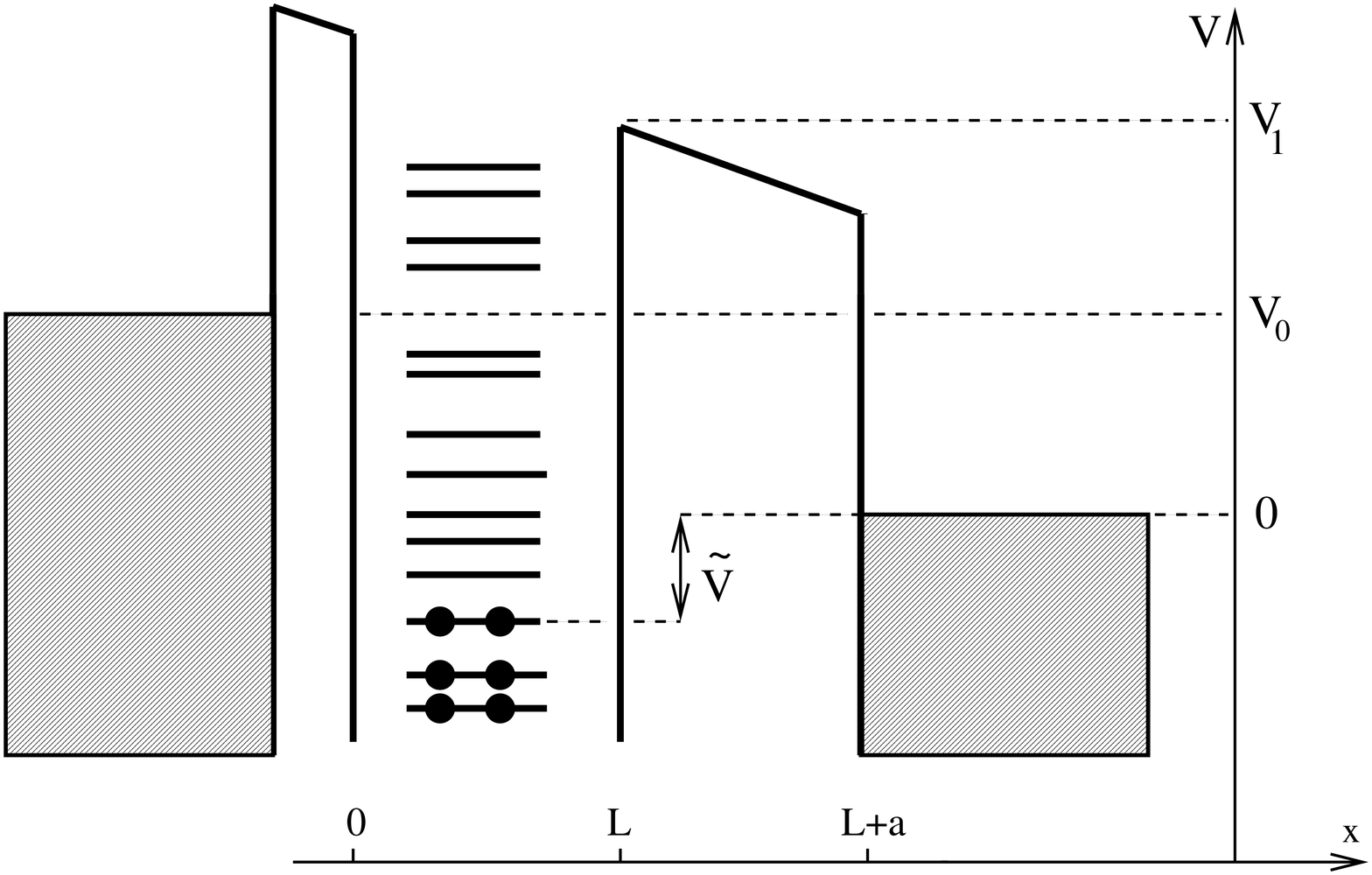} 
\caption{Illustration of the experimental setup we propose in order to
measure an ultrasharp crossover. An applied electric field imposes a bias to
the right. With a left-hand barrier much thinner than the right-hand one and
large electron-electron repulsion, current flows to the right \textit{via}
single-electron processes and the decay rate is determined only by tunneling
and activation across the right-hand barrier. }
\label{experiment}
\end{figure}

We now discuss a possible experimental realization of the above model. In an
experiment, information about decay rates can be extracted from the
temperature dependence of the conductivity of a Coulomb blockaded quantum dot%
\cite{Ashoori} flanked by a double-barrier heterostructure\cite{Datta}. This
setup is illustrated in Fig.~\ref{experiment}. We assume that the left-hand
barrier is much thinner than the right-hand one. In addition, we assume that
strong decoherence hinders resonant tunneling and that electrons in the dot
are properly equilibrated. Finally, our single-electron picture is valid if
the energy of the electron-electron repulsion in the quasi-equilibrium,
calculated \textit{via} the addition of an electron to the quantum dot, is
larger than $\tilde{V}+V_{0}$. Then the transport through the system is
determined by tunneling to the right from the excited states with energies
between $0$ and $V_{0}$. States with energies between $V_{0}$ and $V_{1}$
tunnel back quickly through the thin left-hand barrier and do not contribute
to the current. Physically, charge transfer through this object occurs in
two stages; first, rapid single-electron tunneling from the left-hand
reservoir into the well and, second, electron escape into the right-hand
reservoir after a relatively long dwelling time in the well.

The current $I$ depends upon the temperature according to 
\begin{eqnarray}
I(T) & \propto & \exp \left( -\frac{{\tilde{V}}}{T}\right) \times   
\\
& \times & \int\limits_{0}^{V_{0}}\,dE\exp \left( -\frac{E}{T}\right) \exp
\left( -\frac{2\sqrt{2m\left( V_{1}-E\right) }a}{\hbar }\right) ,\nonumber
\end{eqnarray}%
where we average the decay rates of single-electron states over the
Boltzmann distribution and substitute the sum (analogous to that in Eq. (\ref%
{decay_rate})) by an integral, as the number of metastable levels in the
well is large. Also, we assume that the temperature is small enough to
neglect the smearing of the Fermi-Dirac distributions in the reservoirs. A
saddle-point analysis similar to the one presented above yields a 
kink in the action
at a critical
temperature 
\begin{equation}
T_{\text{c, dot}}=\frac{\hbar V_{0}}{2\sqrt{2m}a\left( \sqrt{V_{1}}-\sqrt{%
V_{1}-V_{0}}\right) },
\end{equation}%
and, to exponential accuracy, the decay rate is given by 
${\Gamma } 
\propto
\exp 
\left ( - {S_{\text{dot}}(T)}/{\hbar } \right )$, with the effective action
\begin{equation}
\frac{S_{\text{dot}}(T)}{\hbar }%
=\left\{ 
\begin{array}{l@{\quad\quad}l}
{\frac{\tilde{V}}{T}+\frac{2\sqrt{2mV_{1}}a}{\hbar }}, & {T<T_{\text{c, dot}%
},} \\ 
{\frac{{\tilde{V}}+{V_{0}}}{T}+\frac{2\sqrt{2m\left( V_{1}-V_{0}\right) }a}{%
\hbar }}, & {T>T_{\text{c, dot}}}.%
\end{array}%
\right.   \label{final_result_new}
\end{equation}%
This formula is valid for $V_1 > V_0$. For $V_1 < V_0$ one should substitute
$V_1$ by $V_0$.  This leads to the disappearance of the quantum correction
for $T > T_{c,\rm dot}$.
Due to the non-vanishing
value of {the} energy gap $\tilde{V}$ between the Fermi levels of the dot
and of the right-hand reservoir, an electron can tunnel only after having
been raised in energy thermally, resulting in a classical contribution to
the action below $T_{\text{c}}$. (This also explains why $V_{1}$, rather
than $V_{1}+\tilde{V}$, appears for ${T<T_{\text{c, dot}}}$.) 
Experimentally, one monitors the singularity in $I(T)$ at $T=T_{\text{c, dot}%
}$, scanned by varying $\tilde{V}$ (through the gate voltage) and $V_{1}$
(through the Fermi energy of the left-hand reservoir). If ${\tilde{V}}%
\rightarrow 0$ and $V_{1}\rightarrow V_{0}$, Eq. (\ref{final_result_new})
reduces to the simpler dependence of Eq.~(\ref{final_result}).

If the electric field $\mathcal{E}$ that biases the system is too large, the
potential is significantly distorted away from a square well. We thus
require $a\ll L$. Otherwise, the action $S(T)$ is not given by Eq. (\ref%
{final_result_new}) and one has to take a non-vanishing electric field into
account. This, however, does not change the conclusion of the existence of a
sharp crossover 
although it becomes less abrupt and may
not be classified as `ultrasharp' for large enough $\mathcal{E}$\cite%
{unpublished}. Another experimental difficulty relates to the fact that real
potentials are smeared compared to square wells; our theory is applicable if
the characteristic size of the smearing is much smaller than the size $a$ of
the barrier. Within the quantum dot itself the potential is difficult to
control; in order to observe an ultrasharp crossover one needs to flank the
dot with a double-barrier structure. It is experimentally possible to create
linear potentials using semiconductor heterostructures\cite{Datta}---a
potentially useful technique if the condition $a\ll L$ cannot be satisfied.
The field $\mathcal{E}$ then can be chosen to yield a resulting rectangular
right-hand barrier. The ultrasharp transition 
originates in the existence of a large enough region $L<x<L+a$ in
which the potential does not vary substantially, while the detailed shape of
the potential within the quantum dot itself is largely irrelevant.

Before concluding, we point out that the metastability condition $S(T)\gg
\hbar $ cannot be satisfied in practice arbitrarily well. If the action $%
S(T) $ is too large, the tunneling time exceeds the duration of experiments,
and no current is detected. Typically, one requires $S(T)/\hbar \alt30$ in
order to observe decay. Because of the finiteness of $S(T)$, the crossover
from quantum to classical behavior is rounded over a narrow region, the
width of which is estimated as follows. As long as the difference between
classical and quantum actions, divided by $\hbar $, is of order 1, \textit{%
i.e.}, as long as 
\begin{equation}
\left\vert \frac{S_{0}}{\hbar }-\frac{V_{0}}{T}\right\vert \alt1,
\end{equation}%
neither of the two processes dominates over the other. Assuming $%
V_{0}/T_{c}\gg 1$ and expanding $1/T$ in $\Delta T=T-T_{c}$, we find that
the crossover is rounded over an interval $\Delta T\simeq T_{c}^{2}/V_{0}\ll
T_{c}$.

In summary, we showed that the decay rate of a metastable electron in a
rectangular well exhibits an ultrasharp transition from quantum to classical
behavior: to exponential accuracy, decay results from quantum tunneling
only, below a critical temperature $T_{\text{c}}$ set by the barrier height,
while above $T_{\text{c}}$, only thermal activation is relevant. Moreover,
we described a semiconductor heterostructure that may be used to check our
theoretical prediction, as well as some of the associated experimental
restrictions.

We thank R.C.~Ashoori, P.W.~Brouwer, C.M.~Marcus, and R.M.~Westervelt for
helpful discussions. This work was supported by the Packard Foundation
(DAG), the \textit{Fonds National Suisse} through a Young Researcher
Fellowship and the Harvard Society of Fellows (RAS).


\begin{thebibliography}{99}
\bibitem{Hanggi} P.~H\"anggi, P.~Talkner, and B.~Borkovec, Rev. Mod. Phys. 
\textbf{62}, 251 (1990).

\bibitem{Blatter} G.~Blatter, M.V.~Feigel'man, V.B.~Geshkenbein,
A.I.~Larkin, and V.M.~Vinokur, Rev. Mod. Phys. \textbf{66}, 1125 (1994).

\bibitem{Coleman} S.~Coleman, Phys. Rev. D \textbf{15}, 2929 (1977);
C.G.~Callan and S.~Coleman, Phys. Rev. D \textbf{16}, 1762 (1977).

\bibitem{Linde} A.D.~Linde, Nucl Phys. B \textbf{216}, 421 (1983), and
references therein.

\bibitem{Rajaraman} R.~Rajaraman, \textit{Solitons and Instantons}
(Elsevier, Amsterdam, 1982).

\bibitem{Affleck} I. Affleck, Phys. Rev. Lett. \textbf{46}, 388 (1981).

\bibitem{Chudnovsky} E.M. Chudnovsky, Phys. Rev. A \textbf{46}, 8011 (1992).

\bibitem{Ch1} E.M.~Chudnovsky and D.A.~Garanin, Phys. Rev. Lett \textbf{79},
4469 (1997).

\bibitem{Gorokhov} D.A.~Gorokhov and G.~Blatter, Phys. Rev. B \textbf{56},
3130 (1997).

\bibitem{MK} J.Q.~Liang, H.J.W.~M\"{u}ller-Kirsten, D.K.~Park, and
F.~Zimmerschied, Phys. Rev. Lett. \textbf{81}, 216 (1998).

\bibitem{Kim} G.H.~Kim, Phys. Rev. B \textbf{67}, 144413 (2003).

\bibitem{RLu} R.~L\"{u}, S.-P.~Kou, J.-L.~Zhu, L.~Chang, and B.-L.~Gu, Phys.
Rev B \textbf{62}, 3346 (2000).

\bibitem{dynamics} The simple inertial dynamics we consider here could be
generalized to include dissipation or a Hall contribution; see Ref. \cite%
{Blatter}.

\bibitem{Larkin} A.I.~Larkin and Yu.N.~Ovchinnikov, Pis'ma Zh. Eksp. Teor.
Fiz. \textbf{37}, 322 (1983) [JETP Lett. \textbf{37}, 382 (1983)].

\bibitem{Kent} L. Bokacheva, A.D. Kent, and M.A. Walters, Phys. Rev. Lett. 
\textbf{85}, 4803 (2000).

\bibitem{Ashoori} For a review, see R.C.~Ashoori, Nature \textbf{379}, 413
(1996).

\bibitem{Datta} C.~Weisbusch and B.~Vinter, \textit{Quantum Semiconductor
Structures} (Academic Press, New York, 1991).

\bibitem{unpublished} D.A.~Gorokhov and R.~A.~da~Silveira, unpublished. 
\end{thebibliography}
\end{document}